\documentstyle{mn}
\title[]{Intrinsic Parameters of GRB990123 from Its Prompt\\
Optical Flash and Afterglow } 

\author[]{X. Y. Wang$^{1}$, Z.G. Dai$^{1,2}$ and T. Lu$^{1,2}$ 
\thanks{E-mail: xywang@nju.edu.cn; daizigao@public1.ptt.js.cn;
tlu@nju.edu.cn} \\
\rm $^1$Department of Astronomy, Nanjing University, Nanjing 210093, 
         P. R. China \\
\rm $^2$LCRHEA, Institute for High-Energy Physics, Chinese Academy of 
         Sciences, Beijing 100039, P. R. China} 
\date{Accepted ........
      Received .......;
      in original form .......}
\pubyear{2000}

\begin{document}

\maketitle 

\begin{abstract}
We have constrained the intrinsic parameters, such as 
the magnetic energy density fraction ($\epsilon_{B}$),
the electron energy density
fraction ($\epsilon_e$),  the initial
Lorentz factor ($\Gamma_0$) and the Lorentz factor of
the reverse external shock ($\Gamma_{rs}$), of GRB990123, 
in terms of the afterglow information
(forward shock model) and the optical flash information (reverse shock model).
Our result shows:  1) the inferred values of $\epsilon_e$ and
$\epsilon_B$  
are consistent with  the suggestion
that they may be  universal parameters, comparing to those inferred for 
GRB970508; 
2) the
reverse external shock may have become relativistic before it passed
through the ejecta shell. 
Other instrinsic parameters of GRB990123, such as energy contained
in the forward shock $E$ and the ambient density $n$
are also determined  and discussed in this paper.
\end{abstract}
\begin{keywords}
gamma-rays: bursts---shock waves---magnetic fields---optical
radiation
\end{keywords}

\section{Introduction}
The current standard model for gamma-ray bursts (GRBs) and their afterglows
is the fireball-plus-shock model (see Piran 1999 for a review).
It involves that a large amount of energy, $E_{0} \sim10^{51-54}$ ergs,
is released within a few seconds in a small volume with negligible
baryonic load, $Mc^2\ll{E_0}$. This leads to a fireball that expands
ultra-relativistically with a Lorentz factor $\Gamma_0\simeq{E_0/Mc^2}>
100$ required to avoid the attenuation of hard $\gamma$-rays due to pair
production
(e.g. Woods $\&$ Loeb 1995; Fenimore, Epstein $\&$ Ho 1993).
A substantial fraction of the kinetic energy of the baryons is transferred
to a non-thermal population of relativistic electrons through Fermi
acceleration in the shock ($\rm{M{\acute{e}}sz{\acute{a}ros}}$ $\&$
Rees 1993). The accelerated electrons cool via synchrotron emission and
inverse Compton scattering in the post-shock magnetic fields and produce
the radiation observed in GRBs and their afterglows (e.g. Katz 1994;
Sari et al. 1996; Vietri 1997; Waxman 1997a; Wijers et al. 1997).
The shock could be either $\it{internal}$ due to collisions between
fireball shells caused by outflow variability (Paczy$\rm\acute{n}$ski
$\&$ Xu 1994; Rees $\&$ $\rm{M{\acute{e}}sz{\acute{a}ros}}$ 1994),
or $\it{external}$ due to the interaction of the fireball with the
surrounding interstellar medium (ISM; $\rm{M{\acute{e}}sz{\acute{a}ros}}$
$\&$ Rees 1993). The radiation from internal shocks can explain
the spectra (Pilla $\&$ Loeb 1998) and the fast irregular variability
of GRBs (Sari $\&$ Piran 1997), while the synchrotron emission from
the external shocks provides a successful model for the broken power
law spectra and the power law decay of afterglow light curves
(e.g. Waxman 1997a,b; Wijers, Rees $\&$ $\rm{M{\acute{e}}sz{\acute{a}ros}}$
1997; Vietri 1997; Dai $\&$ Lu 1998a,b,c; Dai et al. 1999; Wang et al.
1999a,b;
Huang et al. 1998a,b; 1999a,b).

The properties of the synchrotron emission from GRB shocks are determined
by the magnetic field strength, $B$, and the electron energy distribution
behind the shock. Both of them are difficult to estimate from first
principles, and so the following dimensionless parameters are often used
to incorporate modeling uncertainties (Sari et al. 1996),
$\epsilon_B\equiv{\frac{U_B}{e_{th}}}$, $\epsilon_e\equiv{\frac{U_e}
{e_{th}}}$. Here $U_B$ and $U_e$ are the magnetic and electron energy
densities and $e_{th}=nm_{p}c^{2}(\gamma_p-1)$ is the total thermal
energy density behind the shocks, where $m_p$ is the proton mass, $n$
is the proton number density, and $\gamma_p$ is the mean thermal
Lorentz factor of the protons. In spite of these
uncertainties, an important assumption that $\epsilon_B$ and $\epsilon_e$
do not change with time, has been made in the standard external shock
model. Through the computation, Wijers $\&$ Galama (1998; hereafter WG99)
even suggested
that they may be universal parameters, i.e. the same for different
bursts.

The BeppoSAX satellite ushered in 1999 with the discovery of GRB990123
(Heise et al. 1999),
the brightest GRB seen by BeppoSAX to date.
This is a very strong burst. 
An assumption of isotropic
emission and the detection of the source's redshift
$z={1.6004}$ , lead to
a huge energy release about $1.6\times{10^{54}}$ergs (Briggs et al. 1999;
Kulkarni et al. 1999a) in $\gamma$-rays
alone. GRB990123 would have been  amongest the most exciting
GRBs even just on the basis of these facts. Furthermore, ROTSE discovered
a prompt optical flash of 9-th magnitude (Akerlof et al. 1999).
It is the first time that a prompt emission in another wavelength
apart from $\gamma$-rays has been detected from GRB. Such a strong
optical flash was predicted to arise from a reverse external shock propagating
into the relativistic ejecta ($\rm{M{\acute{e}}sz{\acute{a}ros}}$ $\&$
Rees 1997;  Sari
$\&$ Piran 1999a,b, hereafter SP99a,b). This is the so called
``early afterglow". The five last exposures of ROTSE show a power
law decay with a slope of $\sim$2.0,
which can also be explained by the reverse shock model (SP99b;
 $\rm{M\acute{e}sz\acute{a}ros}$ $\&$ Rees 1999).
The usual afterglows in X-ray, optical, IR and radio bands were
also detected after the burst.
They have two distinguishing features: 1)
 the radio emission is unique both due to its very early
appearance and its rapid decline;
2) the temporal decaying behaviour of the $r$-band light
curve after two days steepens from about $t^{-1.1}$ to $t^{-1.8}$
(Kulkarni et al. 1999a; Fruchter et al. 1999; Casrto-Tirado et al. 1999),
 and this steepening
might be due to a jet which has transited from a spherical-like phase
to sideways expansion phase (Rhoads 1999; Sari et al. 1999; Wei $\&$ Lu
1999) or
a dense medium which has slowed down the relativistic expansion of
shock quickly to a non-relativistic one (Dai $\&$ Lu 1999a,b).

Galama et al. (1999)  assumed that the radio emission
is produced by the forward shock as usual and then
reconstructed the radio-to-X-ray afterglow spectrum
on January 24.65 UT.  
However, later work shows that 
the simplest interpretation of this
``radio flare" is that it arises in the reverse shock and that 
such radio emission is an inevitable consequence of the 
prompt bright optical flash seen by ROTSE (Kulkarni et al. 1999b;
Sari $\&$ Piran 1999b).
Kulkarni et al. (1999b) also constrained two key parameters of
the forward shock, the peak flux $F_{\nu_m}$ and the peak
frequency $\nu_m$, to within a factor of two. 
For previous bursts, we have no other 
information apart from
the afterglow to infer the intrinsic parameters of the external shocks.
But now the optical flash of GRB990123 has been fortunately detected, which
enables us to determine another two key parameters, the initial 
Lorentz factor $\Gamma_0$ and the Lorentz factor of the reverse shock
$\Gamma_{rs}$.
 WG99  computed the intrinsic parameters of
GRB970508 and GRB971214 in terms of their {\it afterglow} data, and found
that  $\epsilon_e$ 
is nearly the same for these two bursts,
suggesting it may be a universal parameter.
Granot, Piran $\&$ Sari (1998a,b; hereafter GPS99a,b) modified the set of
equations derived
by WG99, and inferred the electron energy density
fraction and the magnetic energy density
fraction of GRB970508 to be:  $\epsilon_e=0.57$;
$\epsilon_B=8.2\times10^{-3}$.
Here we apply the set of equations of GPS99a  to GRB990123 and
try to determine some intrinsic parameters  based on the information 
of the two aspects of GRB990123---{\it the optical flash and the afterglow}.

The initial Lorentz factor $\Gamma_0$ is also an important physical
parameter of GRBs. It is a crucial ingredient for constraining models
of the source itself, since it specifies how ``clean" the fireball is
as the baryonic load is $M\simeq{E_0/\Gamma_0{c^2}}$. 
Unfortunately, the spectrum of GRBs can provide only a lower limit
to this Lorentz factor ($\Gamma_0 > 100$). Moreover, the current afterglow
observations, which detect radiation from several hours after the burst,
do not provide a verification of the initial extreme relativistic motion.
A possible method to estimate $\Gamma_0$ of GRBs has been suggested by
Sari $\&$ Piran (1999a), based on identifying the ``early afterglow"
peak time. In this paper, the initial Lorentz factor has been inferred
more precisely from the full set of equations describing the reverse
shock region.

In section 2 and 3, we  compute the intrinsic parameters of GRB990123
from its afterglow and  optical flash information. 
In the final section, we give our
conclusions.

\section{ Parameters from the afterglow}

Intrinsic parameters like 
the magnetic energy density fraction $\epsilon_{B}$, electron energy density
fraction $\epsilon_e$, energy in the forward external 
shock $E\equiv{E_{52}\times 10^{52}}$ ergs and  ambient density $n$
can be determined from the afterglow spectrum (GPS99a,b; WG99),
 i. e. if we
 know all three break frequencies (not necessary at the same time)
and the peak flux of the afterglow, we can infer all these
parameters. 

>From the observations of the afterglow of GRB990123,
Kulkarni et al. (1999b) have estimated two key parameters of the 
forward shock: $\nu_m\sim1.1\times10^{13}~{\rm Hz}$  and
$F_{\nu_m}\sim{170}~\mu{\rm Jy}$ at the time $t_{*}=1.25~ {\rm days}$
after the burst.
The cooling frequency $\nu_{c}$ cannot be seen from the
 radio-to-X-ray spectrum obtained by Galama et al (1999). 
 This indicates that $\nu_{c}$ is at or above the
 X-ray frequencies. We need to determine it more precisely. The X-ray
 afterglow, observed 6 hours after the burst, decayed with
 $\alpha_{X}=1.44\pm{0.07}$ ( Heise et al. 1999), while the optical
 afterglow with $\alpha_{r}=1.10\pm{0.03}$ (Kulkarni et al., 1999).
 An X-ray afterglow decay slope steeper by $\frac{1}{4}$
 than an optical decay, which seems to be the case in
this burst, 
 is predicted by Sari, Piran
 and Narayan (1998), if the cooling frequency is between the
 X-rays and the optical.
 So at the time 6 hours after the burst , $4\times{10^{14}}\leq\nu_{c}
 \leq\nu_{X}\sim{(4.4-44)\times{10^{17}}}$Hz. Extrapolating it to
 the time $t_{*}$ as $\nu_{c}\propto{t^{-\frac{1}{2}}}$, we
 get $1.8\times{10^{14}}$Hz$\leq\nu_{c}(t_{*})\leq(2-20)\times{10^{17}}$Hz.
 Another speculative constraint on $\nu_{c}$ is obtained
 from the GRB spectrum itself by Sari $\&$ Piran (1999b), who constrained
 $\nu_{c}\geq2\times{10^{19}}$Hz at the time $t\sim50$ sec.
 Extrapolating it to $t_{*}$, we get $\nu_{c}(t_{*})\geq0.4\times{10^{18}}$.
  Now $\nu_{c}$ is almost determined, and we take the approximate value
$\nu_{c}
  =0.5\times{10^{18}}$Hz, which is in agreement with the estimate of
  Galama et al (1999). 
  In addition, Kulkarni et al. (1999a) have inferred the electron index 
$p$
  (defined as $N(\gamma_e)\propto{\gamma_e}^{-p}$) to be $p=2.44$.
  Now, apart from the self-absorption frequency $\nu_a$, we have all
 other three quantities of the afterglow spectrum
 required to calculate the intrinsic parameters:
 \begin{equation}
 \nu_{m}\sim1.1\times10^{13}{\rm Hz,}~~~~
\nu_{c}\sim0.5\times{10^{18}} {\rm Hz,}~~~~
F_{\nu_m}\sim170~\mu{\rm Jy,}~~~~p\sim2.44                    \\
\end {equation}

Following GPS99b, we adopt the formulas of $\nu_m$ and $F_{\nu_m}$ from 
GPS99a and $\nu_c$ from Sari, Piran $\&$ Narayan (1998):
\begin{equation}
\nu_m\simeq2.9\times10^{15}(1+z)^{1/2}(\frac{p-2}{p-1})^{2}{\epsilon_B}^{1/2}
{\epsilon_e}^2{E_{52}}^{1/2}{t_d}^{-3/2}  {\rm Hz};
\end{equation}
\begin{equation}
F_{\nu_m}\simeq{1.7}\times10^{4}(1+z){\epsilon_B}^{1/2}E_{52}{n_{1}}^{1/2}
(\frac{d_{L}}{10^{28}\rm cm})^{-2} \mu{\rm Jy};
\end{equation}
\begin{equation}
\nu_c\simeq2.7\times10^{12}{\epsilon_B}^{-3/2}{E_{52}}^{-1/2}n^{-1}
{t_d}^{-1/2}(1+z)^{-1/2}  {\rm Hz},
\end{equation}
where $z$ is the redshift of the burst and
$d_L=2c(1+z-\sqrt{1+z})/H_0$ is the luminosity distance.
 
Now we have three equations (2), (3) and (4) with four unknowns:
$\epsilon_e$, $\epsilon_B$, $E_{52}$ and $n$. To solve these equations,
we here assume that the electron density fraction $\epsilon_e$
is the same for different bursts,  just an argument of
WG99, though in which a different set of equations are used.
Since we here use the formulas of GPS99a, we adopt the value of $\epsilon_e$  
from that of GRB970508 inferred  according to the above formulas,
i.e. $\epsilon_e\sim0.57$(GPS99b).

By combining Eqs.(1) and (2)-(4), we get the values of four intrinsic
parameters of the forward shock region:
\begin{equation}
\epsilon_e\sim0.57 ~~~~\epsilon_B\sim3.1\times10^{-3}  ~~~~\\
n\sim0.01~~~~ E_{52}\sim5
\end{equation}

We astonishingly find that the value $\epsilon_B$ inferred is very
close to that of GRB970508 ($\epsilon_B\simeq8.2\times10^{-3}$;
GPS99b), considering the uncertainties in the $\nu_m$ and $F_{\nu_m}$
of a factor of two. This result supports our above adoption of the value
of $\epsilon_e$ and implies that 
 $\epsilon_e$ and $\epsilon_B$
may be universal parameters (i.e. constants for every GRBs), favouring
the argument of WG99.

The ambient density $n$ inferred for GRB990123 is $n\sim0.01$.
This density is on the low side of normal for a disc of galaxy but definitely
higher than expected for a halo, lending a further support to the
notion that bursts occur in gas-rich environment.
The inferred isotropic energy left in the adiabatic forward shock 
is $E\sim5\times10^{52}$ergs, about thirty times less than the isotropic
energy in $\gamma$-rays. This case is very similar to GRB971214 (WG99).
We think, for GRB990123, there are two processes causing
$E_{52}<E_{52,\gamma}$.
One is that there is a radiative evolution phase before the adiabatic phase, 
causing it to emit more of the initial explosion energy and leaving
less for the adiabatic phase.
According to Sari, Piran $\&$ Narayan (1998), we can estimate the reduced 
energy of the forward shock  in the self-similar deceleration stage, i.e.
\begin{equation}
E_{f,52}\sim0.02{\epsilon_B}^{-3/5}{\epsilon_e}^{-3/5}{E_{i,52}}^{4/5}
( {\Gamma_A}/100)^{-4/5}n^{-2/5},
\end{equation}
 where $E_{i,52}$ denotes 
 the  initial
isotropic energy of the forward shock during the self-similar stage
and $E_{f,52}$ denotes the final energy after the radiative phase.
The value of $E_{i,52}$ is difficult to be determined precisely and it may be
less than  $E_{52,\gamma}$ for GRB990123, according to Freedman \& 
Waxman (2000).
If we
use $E_{i,52}\sim E_{52,\gamma}$ ,
 $\epsilon_e\sim0.57$, $\epsilon_B\sim8.2\times10^{-3}$ (instead of 
 $\epsilon_B\sim3.1\times10^{-3}$, since the latter is less  reliable than the
 former, considering the uncertainties in the $\nu_m$ and $F_{\nu_m}$ of a
factor 
 of two for GRB990123.),
$\Gamma_A\sim300$ (see the next section)
 and $n\sim0.01$, we then get $E_{f,52}\sim56$.
The other process is the sideways expansion of the fireball jet (Kulkarni
et al.1999a,b), which can also reduce the energy per solid angle,
hence the isotropic energy $E_{52}$.
Since the opening angle of the jet is $\theta=\theta_0+c_{s}t_{proper}/ct
\sim\theta_0+{\gamma}^{-1}$ (Sari, Piran $\&$ Halpern 1999; Rhoads 1999),
at the time $t=1.25~$days (very near the break time of the jet evolution
$t_b\sim2.1~$days), $\gamma\sim\frac{1}{\theta_0}$ (note that $\gamma
\propto{t^{-3/8}}$), then $\theta\sim2\theta_0$.
 Therefore, the real isotropic energy $E_{52}$ 
left in the late adiabatic forward shock should be  $56(\frac{2\theta_0}
{\theta_0})^{-2}\sim{14}$ , in rough agreement with the above value inferred
from the afterglow spectrum.  An additional possible loss of energy may be
the reverse shock itself if it is radiative.

\section{Parameters from the optical flash}

An optical flash is considered to be produced by the reverse external shock,
which heats up the shell's matter and accelerates its electrons
(SP99b; {\rm M$\acute{e}$sz$\acute{a}$ros} $\&$ Rees 1999). BATSE's
observations triggered ROTSE via BACODINE system (Akerlof et al. 1999).
An 11.82 magnitude optical flash was detected on the first 5
seconds exposure, 22.18 seconds after the onset of the burst. Then the
optical emission peaked in the following 5 seconds exposure, 25 seconds
later, which revealed an 8.95 magnitude signal ($\sim$1Jy). The optical
signal decayed to 10.08 magnitude 25 seconds later and continued to decay
down to 14.53 magnitude in the subsequent three 75 seconds exposures that
took place up to 10 minutes after the burst. The five last exposures depict
a power law decay with a slope $\sim2.0$ (Akerlof et al. 1999; SP99b).
Sari $\&$ Piran (1999b) and {\rm M$\acute{e}$sz$\acute{a}$ros} $\&$ Rees
(1999)
assumed that the ejecta shell follows the Blandford-McKee (1976) self-similar
solution after the reverse shock has passed through it and explained the 
decay of $t^{-2.0}$.

So we assume that at the optical emission peak time ($t=50$~sec) the
reverse shock
had just passed through the ejecta shell. At this time, the Lorentz
factor of the reverse shock $\Gamma_{rs}$ is approximately given by
\begin{equation}
\Gamma_{rs}=\frac{\Gamma_0}{\Gamma_{A}},
\end {equation}
where $\Gamma_0$ is the initial Lorentz factor of the ejecta and $\Gamma_A$
is the Lorentz factor of the ejecta at the optical flash peak time. 
 Then the random minimum Lorentz factor $\gamma_{min}$ of the  
electrons in the reverse shock region is:
\begin{equation}
\gamma_{min}=\frac{m_p}{m_e}\frac{p-2}{p-1}\epsilon_e\frac{\Gamma_0}{\Gamma
_A},
\end {equation}

The formulas of $\nu_m$ at the reverse shock
were given by Sari $\&$ Piran
(1999b). We here add the correction for redshift.
\begin{equation}
\nu_m=1.2\times10^{13}(\frac{\epsilon_e}{0.1})^{2}(\frac{\epsilon_B}
{10^{-3}})^{1/2}(\frac{\Gamma_0}{300})^{2}n^{1/2}(1+z)^{-1} {\rm Hz}
\leq{5}\times10^{14} {\rm Hz}.
\end{equation}

The observed flux at $\nu_m$ can be obtained by assuming that all the
 electrons in the reverse shock region contribute the same average
power per unit frequency $P'_{\nu_m}$ at $\nu_m$, which is given by
$P'_{\nu_m}=\frac{\sqrt{3}e^{3}B'}{m_e{c^2}}$, where 
$B'=\Gamma_{A}c{\sqrt{32\pi{nm_p}\epsilon_B}}$.
Adding one factor of $\Gamma_{A}$ to transform to the observer frame and
accounting for the redshift, we have:
\begin{equation}
F_{\nu_m}=\frac{N_e\Gamma_{A}P'_{\nu_m}(1+z)}{4\pi{d_L}^{2}},
\end {equation}
where $N_e$ is the total number of radiating electrons in the ejecta
shell, and $d_L=2c(1+z-\sqrt{1+z})/H_0$ is the luminosity distance.
Please  note that  $N_e$ here is different from the $N_e$ adopted in
the forward shock region, which is the total number of swept-up
electrons by the forward external shock. We consider $N_e$ here to be
 the total number
of electrons contained in the baryonic load:
\begin{equation}
N_e=\frac{M}{m_p}\simeq\frac{E_{\gamma}}{\Gamma_0{m_p{c^2}}}=1.08\times10^{54}
(\frac{E_\gamma}{1.6\times10^{54}})(\frac{\Gamma_0}{1000})^{-1},
\end {equation}
where $E_{\gamma}$ is the total energy in $\gamma$-rays.
Substituting the expression of $P'_{\nu_m}$ and $z=1.6$ into Eq.(10), we get
\begin{equation}
F_{\nu_m}=0.74(\frac{E_\gamma}{1.6\times10^{54}})(\frac{\Gamma_0}{1000})^{-1
}(\frac{\Gamma_A}
{100})^{2}n^{1/2}(\frac{\epsilon_B}{10^{-3}})^{1/2} {\rm Jy}.
\end{equation}
Please note that this formula always holds whether the ejecta is jet-like
or spherical, because the beaming factor in Eq.(10) and (11) will cancel
out each other in the jet-like case. According to the jump
condition of the shock, the Lorentz
factor of the shocked shell should be approximately equal to that
of the shocked ISM (Piran 1999). The Lorentz factor
of the forward shocked ISM can be obtained from the standard afterglow
model (e. g. Sari, Piran $\&$ Narayan 1998):
\begin{equation}
\Gamma_{A,fs}(t)\simeq6(\frac{E_{52}}{n})^{1/8}(\frac{t_d}{1+z})^{-3/8}
\end {equation}
For $E_{52}\sim5$ and $n\sim0.01$, we get
\begin {equation}
\Gamma_{A,rs}(50s)=\Gamma_{A,fs}(50s)=300
\end {equation}
Taking the above inferred value $\epsilon_B\sim3.1\times10^{-3}$, from
the Eqs.(9), (12), and (14)  with the conditions:
$\nu_m\leq5\times10^{14}$Hz and $F_{\nu_m}\sim1$Jy,
we finally get:
\begin{equation}
\Gamma_A=300,~~ \Gamma_0=1200,~~ \epsilon_e\leq0.6.
\end{equation}

Please note that here the  value $\epsilon_e\leq0.6$  inferred from the
optical flash
data is consistent with that inferred independently from the afterglow
information. On the other hand, if we substitute the value $\epsilon_e=0.57$
into Eq.(9), we find that the peak frequency of the reverse shock $\nu_m$
is almost located at the optical band.         
In addition, our inferred initial Lorentz factor $\Gamma_0$ is six times
larger than
that obtained by Sari $\&$ Piran (1999b), who have used the ambient density
$n$
of GRB970508.
Consequently, at the time the reverse shock has just passed
through the ejecta shell, its Lorentz factor was $\Gamma_{rs}=\frac{
\Gamma_0}{\Gamma_A}\sim{4}$. This indicates that the reverse shock had
become relativistic before it crossed the entire shell. This
result is also different from that obtained by Sari $\&$ Piran (1999b), who
found the Lorentz factor of the reverse shock of GRB990123 was only 
near one. However we argue that our result is
reasonable according to the criterion presented by Sari $\&$ Piran (1995)
(also see Kobayashi et al. 1998).
They defined a dimensionless parameter $\xi$ constructed from $l$, $\Delta$
and $\Gamma_0$:
\begin{equation}
\xi\equiv(l/\Delta)^{1/2}{\Gamma_0}^{-4/3},
\end{equation}
where $l=(\frac{E}{nm_p{c^2}})^{1/3}$ is the Sedov length,
$\Delta=c\delta{T}$ is the width of the shell ($\delta{T}$ is the duration
of GRB) and $\Gamma_0$ is the initial Lorentz factor of the ejecta. If
$\xi<1$, the reverse shock becomes relativistic before it crosses the shell;
otherwise ($\xi>1$), the reverse shock remains Newtonian or at best
mildly relativistic during the whole energy extraction process. For
GRB990123, we find $\xi\sim0.3<{1}$. So the reverse shock of GRB990123
had become relativistic before it crossed the shell, consistent with
our calculated result.

\section{ Conclusions and Discussions }
We have constrained some intrinsic parameters, such as 
the magnetic energy density fraction ($\epsilon_{B}$),
the electron energy density
fraction ($\epsilon_e$), the isotropic energy in the adiabatic forward shock
$E_{52}$ and the ambient density $n$. Due to the lack of the value
of the self-absorption frequency $\nu_a$, we made an assumption that
$\epsilon_e$ of GRB990123 is the same as that of GRB970508, 
then astonishingly find that the inferred value of $\epsilon_B$
is also nearly equal to that of GRB970508. This result favours
the argument proposed by WG99 that the magnetic energy fraction
and the electron density fraction may be universal parameters.

Another two important intrinsic parameters of GRB990123 are also inferred from
the optical flash information: the initial Lorentz factor $\Gamma_0$
and the Lorentz factor $\Gamma_A$ at the prompt optical emission peak 
time of the ejecta. They are: $\Gamma_0=1200$,
$\Gamma_A=300$. Our inferred value of the $\Gamma_0$ is six times
larger than that obtained by
Sari $\&$ Piran (1999b), who used the ambient density $n$ 
inferred for GRB970508.
A larger initial Lorentz factor is reasonable in
consideration of the huge energy of this burst. The Lorentz factor
of the reverse shock at the optical flash peak time is $\Gamma_{rs}\sim
\frac{\Gamma_0}{\Gamma_A}\sim{4}$, which shows that the
reverse shock had become
relativistic rather than mildly relativistic before it crossed
the entire ejecta shell. This result is in agreement with the criterion
presented by Sari $\&$ Piran (1995) to judge the RRS case or NRS case.

Prompt optical flash has added another dimension to GRB astronomy.
Prompt observations in the optical band during and immediately
after GRB may provide more and more events of optical flash in the near
future, and they will enable us to make more detailed analyses,
make more precise determination of intrinsic parameters and test
the reverse---forward external shock model.

\section*{Acknowledgments}
 We would like to thank the referee for his valuable suggestions
and improvement on our manuscript. X.Y. Wang also thank Dr. D. M. Wei for
helpful discussion.
This work was supported by the National Natural Science Foundation
of China under grants 19773007 and 19825109 and the foundation
of Ministry Education of China.

\end{document}